\documentclass[12pt,a4paper]{article}
\usepackage[utf8]{inputenc}
\usepackage{amsmath}
\usepackage{amsfonts}
\usepackage{amssymb}
\usepackage{graphicx}
\usepackage{epsfig}
\usepackage[left=3.5cm,right=2.2cm,top=3.5cm,bottom=3cm]{geometry}
\usepackage{sectsty}
\sectionfont{\fontsize{12}{11}\selectfont}
\usepackage[labelfont = bf, font={footnotesize,}]{caption}
\newcommand{\cor }{{correlations }}
\newcommand{\fll }{{fluctuations }}
\newcommand{\ring }{{ring-like }}
\newcommand{\jet }{{jet-like }}
\newcommand{\et }{{\(\eta\) }}

\newcommand{\urq }{{\footnotesize URQMD }}
\newcommand{\ipm }{{\footnotesize IPM }}
\newcommand{\qgp }{{\footnotesize QGP }}
\newcommand{\qgpp}{{\footnotesize QGP}}

\newcommand{\src}{{\footnotesize SRC }}
\newcommand{\srcc}{{\footnotesize SRC}}
\newcommand{\lrc}{{\footnotesize LRC }}
\newcommand{\lrcc}{{\footnotesize LRC}}
\newcommand{\fb}{{\footnotesize FB }}
\newcommand{\ff}{{\footnotesize F }}
\newcommand{\bb}{{\footnotesize B }}

\begin{document}

\begin{center}

\textbf{\large Event-by-event multiplicity fluctuations and correlations in ring-like and jet-like events in $^{197}$Au-AgBr collisions at 11.6A GeV/c}\\
\vspace{5mm}
{\bf Bushra Ali\footnote{Bushra.Ali@cern.ch}, Sweta Singh, Anuj Chandra and Shakeel Ahmad} \\
\vspace{3mm}
{\small Department of Physics, Aligarh Muslim University\\
Aligarh 202002, India}

\end{center}

\vspace{1.2cm}
{\footnotesize \noindent \textbf{Abstract:} Event-by-event (ebe) multiplicity \fll and \cor amongst the charged particles emitted in the forward-backward symmetric pseudorapidity (\et) windows of varying widths and positions are investigated by analyzing the experimental data on $^{197}$Au-AgBr collisions at 11.6A GeV/c. The findings are compared with the predictions of relativistic transport model, \urq and independent particle emission (or mixed event) model. It is observed that the \fll in ebe mean pseudorapidity values and those reflected from the \fll strength measure, \(\Phi\) are relatively higher as compared to those expected from the statistically independent particle emission model. The study of the variance, \(\sigma_c^2\) of a suitably defined forward-backward asymmetry variable \(C\) as a function of \et window width and position indicates the presence of strong short-range correlations, which might arise due to isotropic decay of cluster-like objects either in forward or backward \et region. Furthermore, analyses of events having \ring and \jet substructures, carried out separately, suggest that the major contribution to the observed \fll in the data sample are due to ring-like events, while the contributions from the \jet events appear to be rather small. The observed difference in the behavior of correlation strengths from the two types of events might be due to the enhanced emission of Cherenkov gluons, giving rise to the \ring substructure. The mixed event analysis further confirms that the observed \fll are the distinct feature of the data, which disappear after event mixing.}
\vspace{1mm}

\newpage

\noindent {\bf 1. Introduction} \\[-4mm]

\noindent Investigations involving \fll and \cor amongst the produced particles in relativistic nucleus-nucleus (AA) collisions are expected to help check the suggestion whether the \fll of a thermal system are related to various susceptibilities and might serve as an indicator for the possible phase transition\cite{1,2}. Furthermore, the presence of large event-by-event (ebe) fluctuations, if observed,  might be taken as a signal for the \qgp formation\cite{3,4,5,6}. In AA collisions, if the system undergoes a phase transition from hadronic matter to \qgpp, the degrees of freedom in the two phases would be quite different\cite{6}. Due to this difference, \cor and \fll of thermodynamical quantities and(or) the densities of the produced charged particles in a phase space may change, apparently lacking a definite pattern. At {\footnotesize{SPS}} energies, results based on the studies of \fll and \cor indicate that various hadronic observables produced in $^{208}$Pb$^{208}$Pb collisions exhibit quantitative changes in their energy dependence in the {\footnotesize{SPS}} energy range\cite{1,7,8}. Comparing these findings with the predictions of statistical and (or) hadronic transport models, it has been reported that the experimental findings in AA collisions at {\footnotesize{SPS}} energies are consistent with the expected signals of the onset of a phase transition in AA collisions at these energies\cite{1,7,8,9}. It has also been suggested that by studying the deviations of the distributions of ebe mean \(p_T\) (M\(p_T\)) and (or) mean \(\eta\) (M\(\eta\)) from the random distribution expected from {\footnotesize{IPM}}, one may get meaningful information about the randomization and thermalization of the events produced in AA collisions\cite{9,10}. Furthermore, a \fll measure \(\Phi\) has been proposed by Gazdzicki and Mrowczynski\cite{9}, which vanishes in the case of independent emission of particles from a single source. However, if AA collisions are regarded as the independent superposition of nucleon-nucleon (\(nn\)) collisions, \(\Phi\) values should be independent of \(nn\) sub processes and compare well with those obtained for \(nn\) collisions\cite{11}. Moreover, correlations amongst the charged particles produced in various \et bins are, therefore, considered as a powerful tool for understanding the underlying mechanisms of multiparticle production\cite{12,13,14,15}. These correlations, in general, are of two types: the short-range correlations (\srcc) and the long-range \cor (\lrcc)\cite{16,17,18,19,20,21}. Particles having large transverse momentum (\(p_T\)) values are produced through harder perterbative processes and are strongly correlated within short \et distances (\src)\cite{17,22}. On the other hand, particles with lower \(p_T\) values are  produced via soft processes and are believed  to be weakly correlated over rather longer \et range(\lrc)\cite{17,23}. Strong \src have been observed around mid-rapidity in several \(pp\) 
and \(\overline pp\) experiments\cite{16,18,19,20,24}. Findings of these experiments suggest the correlation strength increases with beam energy. However, during the past one and half decades, results based on experiments\cite{14,15,25,26,27,28,29,30,31} and model based studies\cite{12,16,21,31,32,33,34,35,36,37,38,39} exhibit the presence of long-range correlations, which has been understood in terms of multiparton interactions\cite{40} as predicted by the dual parton model\cite{41,42,43,44}. Color glass condensate ({\footnotesize{CGC}}) picture also suggests that the correlations produced at the early stages of the collision may spread across a wide range of pseudorapidity values\cite{16,42,45,46,47,48,49,50}.\\

\noindent After the availability of the data from {\footnotesize{RHIC}} and {\footnotesize{LHC}}, interest in the investigations involving the particle correlations has considerably increased. It is because of the idea that modifications in the cluster characteristics and (or) shortening  in the correlation length in the \et space, if observed at these energies, may be taken as a signal for the \qgp formation\cite{12,13,51}. Although, numerous attempts have been made to study forward-backward (\fb)  multiplicity \cor at {\footnotesize{RHIC}} and {\footnotesize{LHC}} energies, it is however, essential to identify some baseline contribution to the experimentally observed contribution to the experimentally observed \cor which do not depend on new physics, for example, formation of some exotic states, like {\footnotesize{DCC}} or {\footnotesize{QGP}}. Investigations carried out so far also do not give a satisfactory explanation concerning the correlations between the particles produced in \ring  and \jet events\cite{52}. An attempt is, therefore, made to study the \fb \cor in the \ring and \jet events produced in the interactions of $^{197}$Au beam with AgBr nuclei at 11.6A GeV/c. Such a study may also serve as a reference to the future heavy-ion experiment at {\footnotesize{FAIR}} energies. The findings are also compared with the predictions of relativistic transport model, {\footnotesize{URQMD}}. \\
\\

\noindent {\bf 2. Identification of ring-like and jet-like events} \\[-4mm]

\noindent The study of azimuth distributions of relativistic charged particles produced within a narrow \et bin reveals that there are two different kinds of substructures, referred to as the \ring and \jet substructures\cite{52,53,54,55}. If many particles are produced within a narrow \et space but spread over the entire azimuth, \ring substructure would emerge. However, if particles are produced within a narrow \et and \(\phi\) regions, \jet substructure will occur\cite{52,54,55,56}. A schematic diagram of \ring and \jet substructures are displayed in Fig.1. The \ring substructures are expected to occur due to Cherenkov gluons. Each gluon gives rise to a jet. These jets should create a ring with 'jetty' spots or 'jetty' substructure in azimuthal plane which is perpendicular to the primary parton orientation. A large number of Cherenkov gluons are produced at very high energies and may form a ring in a single event. It has also been suggested\cite{52,55,56,57,58,59} that the coherent collective effects in hadronic matter may result in the so-called \ring events. Although the \ring and \jet events do not exhibit significant deviations as expected from its stochastic nature, it may be emphasized here that the studies involving both types of substructures do not give full information of the processes involved and therefore, the two types of substructures should be addressed separately\cite{55}.\\

\begin{figure}[bh]
\centerline{\psfig{file=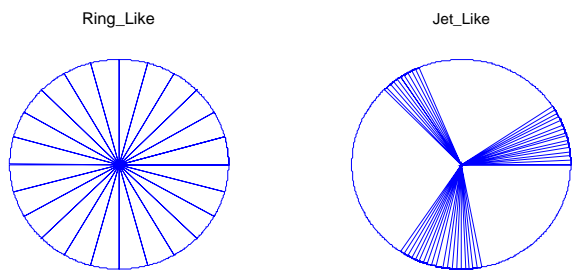,scale=0.45}}
\caption{Schematic diagram of ring-like and jet-like events.}
\end{figure}

\noindent The \ring and \jet events are separated by employing the method proposed by Adamovich et al\cite{53}. According to this approach, a fixed number of \(n_d\) particles is taken and then this \(n_d\) tuple of particles along the \(\eta\)-axis are considered as a group characterized by \(\Delta\eta_c\). The particle density in this \et range is \(\rho_c = n_d/\Delta\eta_c\). Since the multiplicity of particles in each sub-group does not depend on density, therefore, it can be compared with each other. For a given sub-group, the azimuthal structure is to be parameterized in such a way that the larger values of parameter refer to one type of substructure, while the smaller values represent the other one. The following two sums have been suggested\cite{60} for these parameters, namely,

\begin{eqnarray}
 S_1 = -\sum ln(\Delta\Phi_i)
\end{eqnarray}
and
\begin{eqnarray}
 S_2 = \sum ln(\Delta\Phi_i)^2
\end{eqnarray}

\noindent where, \(\Delta\Phi_i\) is the azimuthal difference between two neighboring particles in the group. For the simplicity sake, \(\Delta\Phi_i\) is counted in units of full revolution. This gives:

\begin{eqnarray}
 \sum(\Delta\Phi_i) = 1
\end{eqnarray}

\noindent Both these parameters will be large {\footnotesize{\( (S_1 \rightarrow \infty, S_2 \rightarrow 1)\)}} for \jet substructures and small {\footnotesize{\( (S_1 \rightarrow n_d ln n_d, S_2 \rightarrow 1/n_d)\)}} for \ring substructures. Although, the two parameters, \(S_1\) and \(S_2\) have similar features, \(S_1\) is sensitive to the smallest gap \(\Delta\Phi\), whereas the major contribution to \(S_2\) comes from the largest gap or void in the group\cite{53}. For the present study, the \ring and \jet events are sorted out by evaluating \(\frac{S_2}{<S_2>}\) on ebe basis. If, in an event, \(\frac{S_2}{<S_2>} \leq 1\), it corresponds to a \ring , while the event with \(\frac{S_2}{<S_2>} \geq 1\) is categorized as the \jet event. \\
\\

\noindent {\bf 3. Details of the data} \\[-4mm]

\noindent A sample comprising of 577 events produced in the interactions of 11.6A GeV/c $^{197}$Au beam with AgBr nuclei in nuclear emulsion has been used. This event sample is taken from the series of experiments carried out by {\footnotesize{EMU}}01 collaboration\cite{61,62,63,64}. All the relevant details about the data, like event selection, classification of tracks, extraction of AgBr group of events, methods of measurements, etc., may be found elsewhere\cite{13,21,61,62,63,64}. It should be emphasized that the conventional emulsion technique has two main advantages over the other detectors: (i) its 4\(\pi\) solid angle coverage and (ii) data are free from biases due to full phase space coverage. In the case of other detectors, only a fraction of charged particles are recorded due to the limited acceptance cone. This not only reduces the charged particle multiplicity but may also distort some of the event characteristics, such as particle density fluctuations\cite{11}.\\
\begin{center}
\begin{footnotesize}
\begin{table}
    \begin{tabular}{|c|c|c|c|c|}    \hline
    Event Type  &  \(\langle M_\eta\rangle\)   & \(\sigma_{M_\eta}\)  & \(\omega_\eta\)  &  d\\
    \hline
    Expt.  &  2.480\(\pm\)0.01    &     0.433\(\pm\)0.013   &  0.175\(\pm\)0.005   & 0.043\\
           & (2.441\(\pm\)0.013)  &   (0.321 \(\pm\)0.010)  & (0.132 \(\pm\)0.004) & \\
	\hline

    URQMD & 2.453 \(\pm\) 0.006    & 0.142 \(\pm\) 0.004  & 0.058 \(\pm\) 0.002  & 0.145 \\
          & (2.473 \(\pm\) 0.004)  &(0.107 \(\pm\) 0.003) &(0.043 \(\pm\) 0.001) &\\
	\hline

    Ring-like Event & 2.424 \(\pm\) 0.019 & 0.383 \(\pm\) 0.014  & 0.158 \(\pm\) 0.006  & 0.009\\
                    & (2.391 \(\pm\) 0.018) & (0.356 \(\pm\) 0.013) &  (0.149 \(\pm\) 0.005) &\\
 	\hline

    Jet-like Event & 2.923 \(\pm\) 0.037 & 0.506 \(\pm\) 0.026  & 0.173 \(\pm\) 0.009   & 0.054\\
                   &  (2.913 \(\pm\) 0.025)  &  (0.346 \(\pm\) 0.0018) & (0.118 \(\pm\) 0.006)  &\\
    \hline
    \end{tabular} 
    \caption{Values of \(\langle M_\eta\rangle\), \(\sigma_{M_\eta}\), \(\omega_\eta\) and d for various categouries of events.}
\end{table}
\end{footnotesize} 
\end{center}

\noindent{\bf 4. Definition of observables} \\[-4mm]

\noindent Relativistic charged particles produced in AA collisions are believed to be highly correlated over the large region of pseudorapidity. This raises the question of underlying structure of the single particle distribution\cite{65}. \fb multiplicity \cor studies are considered to be a step forward in this direction. Such \cor are searched for by comparing the event-by-event integrated multiplicities of charged particles emitted in \ff and \bb regions of pseudorapidity. This is done by counting the number of charged particles in a \et window of width \(\Delta\eta\) placed in the forward \et region (\(\eta > \eta_c\)) and a similar \et window placed in the backward \et region (\(\eta < \eta_c\)); \(\eta_c\) being the center of symmetry of \et distribution. These two \et windows are chosen  in such a way that they are symmetric with respect to \(\eta_c\), i.e., centered at (\(\eta_c+\Delta\eta/2\)) and (\(\eta_c-\Delta\eta/2\)) respectively. If \(N_F\) and \(N_B\) respectively denote the event multiplicities in \ff and \bb \et windows, one can obtain the event-wise observable as, \(C = (N_F - N_B)/\sqrt(N_F+N_B)\). The variance of C for a set of events with nominally similar characteristics, is given by\cite{13,52,65}:

\begin{eqnarray}
 \sigma_c^2 = \frac{D_{FF} + D_{BB} - D_{FB}}{\langle N_F + N_B\rangle}
\end{eqnarray}

\noindent where {\footnotesize{\(D_{FF} = \langle N_F^2\rangle - \langle N_F\rangle ^2\)}} and {\footnotesize{\(D_{BB} = \langle N_B^2\rangle - \langle N_B\rangle^2\)}} are respectively the variance in \ff and \bb regions, whereas {\footnotesize{\(D_{FB} = \langle N_FN_B\rangle - \langle N_F\rangle\langle N_B\rangle\)}} denotes the covariance; the quantities with \(\langle ...\rangle\) represent the event averaged values. A comparison of bins covering similar kinematical regions envisages, the mean value of C to be zero\cite{13,66}. However, if multiplicity fluctuations are only of statistical nature, then as per to the binomial partitioning of \(N_F + N_B\) the value of \(\sigma_c^2\)  will be $\sim$ 1. Thus, any deviation of \(\sigma_c^2\) from unity might indicate the presence of fluctuations of some dynamical origin\cite{66,67,68,69}.\\
\\

\noindent {\bf 5. Results and Discussions} \\[-4mm]

\noindent Mean values of \et on ebe basis are computed as:

\begin{eqnarray}
	M_{\eta} = <\eta> = \frac{1}{N}\sum \eta_i 
\end{eqnarray}

\noindent where, \(N_{ch}\) is the multiplicity of relativistic charged particles in an event  and having their \et values in the range, \(\eta_c \pm 2.0\). In order to compare the \(M_\eta\) values with the predictions of \urq and mixed event models, the event samples corresponding to these models are simulated and analyzed. \urq events are simulated by using the code, urqmd-3.4\cite{70,71}. The number of events in the simulated sample is kept equal to that in the real data. The events are generated by considering the percentage of interactions which occur in the collisions of projectile with various targets in emulsion\cite{13,72}. The value of impact parameter is so set that the mean multiplicity of relativistic charged particles becomes nearly equal to one obtained for the experimental data.\\

\noindent Mixed event model is defined by the Monte Carlo procedure frequently used to create a sample of artificial events in which \fll and \cor present in the original events are partially destroyed\cite{73}. The original and mixed events are then analyzed in parallel and the findings are compared to extract the magnitude of of signals of interest, which by construction, might be present in the original event but absent in the mixed one. The mixed event method is popular particularly, in studies of resonance decays, particle \cor and ebe fluctuations\cite{73}. There are several versions of the mixed event model. The one, where the limit of an infinite number of original and mixed events gives results identical to Independent Particle Model {\footnotesize(IPM)}, is adopted here. The procedure of generating a mixed event is as follows\cite{73}.

\begin{itemize}
	\item[(i)] An event with multiplicity \(N_{ch}\) is selected from the original event sample.
	\item[(ii)] \(N_{ch}\) particles for the mixed events are drawn randomly and clubed together to have a set of mixed events.
\end{itemize}

\noindent The procedure is then repeated to generate the next mixed event and so on. In  the limit of infinite number of original events, the probability of having two particles from the same event in a single mixed event is zero.\\

\noindent 
Comparison of \(M_\eta\) distributions for the real and \urq data sets with their corresponding mixed events are displayed in Fig.2. Plots in the bottom panel show the distributions of  difference of the data and mixed events. It may be noted that the distributions for the real data do have  long tails as compared to those due to mixed events, indicating the presence of fluctuations other than the statistical ones. Distributions for the \urq events too exhibit similar trends but with somewhat smaller magnitudes as compared to the experimental data. Mean values of \(M_\eta\) and dispersions of \(M_\eta\) distributions are presented in Table~1. Values given within brackets are due to the mixed events. It is noted from the table that \(\langle M_\eta \rangle\) for the real and \urq events are nearly equal but the values of dispersions reflect the distributions for the \urq events somewhat narrower than those obtained from the experimental data. This is noticeable from the figure too. It has been observed\cite{1} that \(M_\eta\) distributions for $^{16}$O-AgBr collisions at 14.5A, 60A and 200A GeV/c and $^{32}$S-AgBr collisions at 200A GeV/c become narrower and shift towards higher values of \(M_\eta\) with increasing beam energy and system size.\\

\begin{figure}[bh]
\centerline{\psfig{file=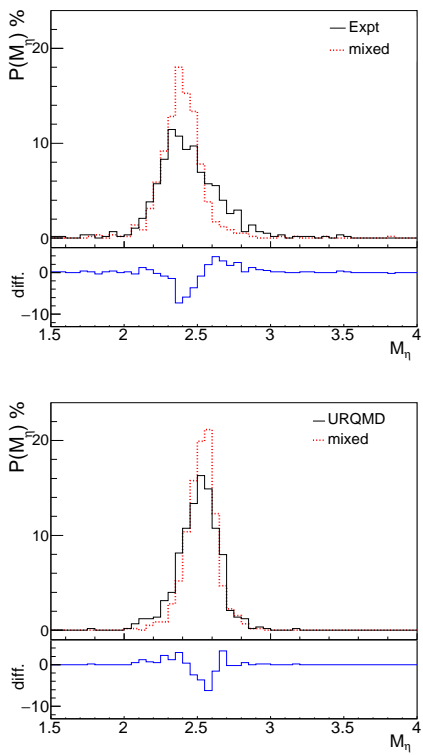,scale=0.4}}
\caption{\(M_\eta\) distributions for the experimental (top) and \urq (bottom) compared with the mixed events.}
\end{figure}

\noindent In order to quantify the deviation of \fll from the one expected from statistically independent particle emission, the magnitude of fluctuations, \(\omega_\eta\) in the quantity \(M_\eta\) is defined as\cite{74}:

\begin{eqnarray}
	\omega_\eta = \frac{\langle M_\eta^2 \rangle - \langle M_\eta \rangle^2}{\langle M_\eta \rangle} = \frac{\sigma_{{M_\eta}}^2} {\langle M_\eta \rangle}
\end{eqnarray}

\noindent and the difference, \(d\) in the values due to the data and mixed events as

\begin{eqnarray}
	d = \omega_\eta(data) - \omega_\eta(mixed)
\end{eqnarray}

\noindent gives the difference in \fll from the random baseline. A value, \(d > 0\), for a given data set would indicate the presence of correlations, like Bose-Einstein correlation\cite{1,10}. It is interesting to note that the values of \(d\) obtained in the present study are positive for both real and \urq data sets. Values of \(d\) for \urq are however, much smaller as compared to the real data.\\

\begin{figure}[th]
\centerline{\psfig{file=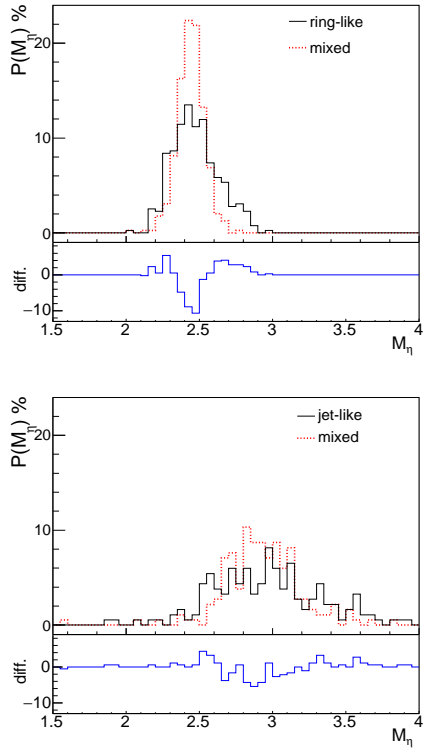,scale=0.4}}
\caption{Comparison of \(M_\eta\) distributions for the \ring and \jet events with their corresponding mixed events.}
\end{figure}

\noindent Fig.3 shows the distributions of \(M_\eta\) for the ring-like and jet-like events compared with the distributions due to mixed events. It is observed in the figure that \(M_\eta\) distribution for jet-like events are much wider as compared to the one with ring-like events. The values of \(\langle M_\eta \rangle\) and \(\sigma_{M_\eta}\) for jet-like events, as are noticed in Table~1, are much higher than those for the ring-like events. However, a comparison of the distributions indicate that \(M_\eta\) distribution for jet like events nearly match with the one due to the mixed events, whereas, in the case of ring-like events, \(M_\eta\) distribution is more populated in the higher \(M_\eta\) region as compared to that for the mixed events.  These observation,therefore, suggest that the dynamical \fll of significant magnitudes are present in the ring-like events, whereas in the jet-like events, such \fll seem to be very small or nearly absent.\\

\noindent It has been argued\cite{1,75} that a comparison of \(M_\eta\) distributions for the data and mixed events may not fully account for the rare dynamical \fll of large magnitude. In order to quantify and examine the deviations of \(M_\eta\) distributions from the baseline, various  other methods have been put forward\cite{9,11,75,76}. The one suggested by Gazdzicki and Mrowezynski\cite{9} is used for the present analysis. According to this method, ebe \fll of observables defined as a sum of particle's kinematical variables, such as, \(\eta\), \(p_T\), etc., where the sum runs over all particles produced in an event within the applied kinematical cuts. By studying the second moment of the distributions of such variables, the degree of randomization and thermalization characteristics of events produced in AA collisions may be evaluated\cite{1,11}. As described in refs.9 and 11, for each particle of an event, a quantity,
 
\begin{eqnarray}
	z_i = \eta_i - \langle \eta \rangle
\end{eqnarray}

\noindent is defined. \(\langle \eta \rangle\) being the mean value of \et distribution of the  entire sample. This gives a variable,

\begin{eqnarray}
	Z = \sum_{i=1}^{N_{ch}} z_i
\end{eqnarray}

\noindent using variables, the measure \(\Phi_\eta\) is calculated as:

\begin{eqnarray}
	\Phi_\eta = \sqrt{\frac{\langle Z^2 \rangle}{\langle N_{ch} \rangle}} - \sqrt{\langle z^2 \rangle}
\end{eqnarray}

\begin{figure}[th]
\centerline{\psfig{file=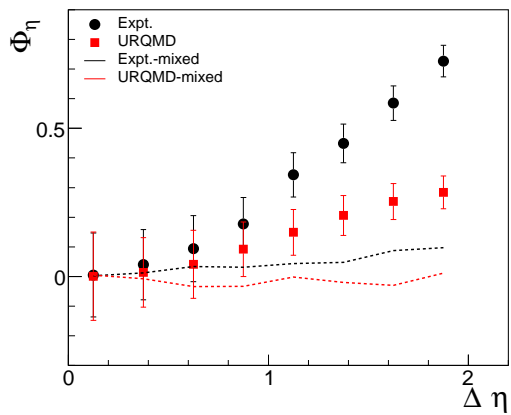,height=6cm}}
\caption{\(\Phi\eta\) Dependence on the width of \et windows, \(\Delta\eta\) for the real and \urq events. The lines represent the results from the mixed events.}
\end{figure}

\noindent where, \(\langle z^2 \rangle\) is the second moment of the inclusive \(z\) distribution. \(\Phi_\eta\) would, thus, quantify the degree of \fll in the values of mean \et from event to event. The value of \(\Phi_\eta\) will be zero for independent emission of particles from a single source. If AA collisions are regarded as the incoherent superposition of multiple independent \(nn\) collisions, the value of \(\Phi_\eta\) should be equal to that for \(nn\) collisions\cite{1,11}. In our earlier work\cite{1}, it has been observed that \(\Phi_\eta\) varies linearly with \(lnE_{beam}\) for $^{16}$O-AgBr collisions at 14.6A, 60A and 200A GeV/c and  $^{32}$S-AgBr collisions at 200A GeV/c, while \(pp\) data at 200 GeV/c give much smaller value of  \(\Phi_\eta\) than the one expected from its observed linear behavior. For the mixed events corresponding to all these data sets,  \(\Phi_\eta\) acquires values \(\sim\) 0, as expected.\\

\begin{figure}[th]
\centerline{\psfig{file=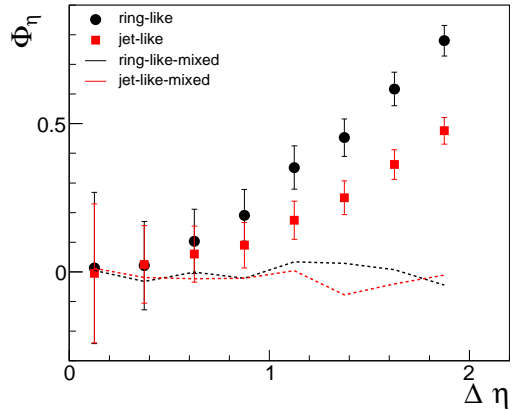,height=6cm}}
\caption{Variations of \(\Phi\eta\) with  \(\Delta\eta\) for the \ring and \jet  events. The lines represent the results from the mixed events.}
\end{figure}

\noindent Shown in Fig.4 are the variations of  \(\Phi_\eta\) with \(\Delta\eta\) for the experimental and \urq events along with the mixed events. It is observed that for \(\Delta\eta\) = 0.25,  \(\Phi_\eta\) = 0 for both experimental and \urq data and then it rises with increasing \(\Delta\eta\). It is, however, noted that experimental data show much faster rise as compared to that of \urq events. The values of  \(\Phi_\eta\) for the mixed events are \(\sim\) 0, irrespective of large or small \(\Delta\eta\) values. Fig.5  shows the dependence of  \(\Phi_\eta\) on \(\Delta\eta\) for the ring-like and \jet events and their corresponding mixed events. It is evident from the figure that with increasing \(\Delta\eta\),  \(\Phi_\eta\) grows rather quickly for \ring as compared to those observed for \jet events. This indicates that \fll present in the \ring events are larger than those present in \jet events.\\

\noindent Multiplicity distributions have been reported to have different patterns of variation in different \et regions\cite{77,78} and exhibit large \fll in wider \et windows\cite{21,79}. In order to study the \fb \cor in \et regions of varying widths, two small windows of width \(\Delta\eta\) = 0.25 are considered and placed symmetrically with respect to \(\eta_c\) such that the separation between the centers of the two windows is \(\Delta\eta\) = 0.25. The charged particles lying in the intervals, \(\eta_c < \eta < \eta_c+\delta\eta\) and \(\eta_c > \eta > \eta_c+\delta\eta\) are counted as \(N_F\) and \(N_B\) and the value of \(\sigma_c^2\) for this \(\Delta\eta\) is evaluated. The window width is then increased in steps of 0.25 until the region \(\eta_c \pm \) 2.0 is covered. Variations of \(\sigma_c^2\) for the experimental and corresponding \urq events are plotted in Figs.6. It is observed in the figure that \(\sigma_c^2\) monotonically increases with increasing widths of \et windows. The increase in \(\sigma_c^2\) values for the real data is noticed to be rather faster than those for \urq events. Variations of \(\sigma_c^2\) with \(\Delta\eta\) for the mixed event sets corresponding to real and \urq events are shown in the right panel of Fig.6. It is interesting to note that \(\sigma_c^2\) for mixed events, acquire values of \(\sim\) 1 throughout. However, for the real and \urq events its values are \(\sim\) 1 for \(\Delta\eta\) = 0.25 and increase with increasing \(\Delta\eta\). These observed dependence of \(\sigma_c^2\) on \(\Delta\eta\) may be explained due to formation or presence context of clusters: With increasing \(\Delta\eta\), the probability that more than one particle resulting from a single cluster may fall in either \ff or \bb region increases.\\ 
\begin{figure}[th]
\centerline{\psfig{file=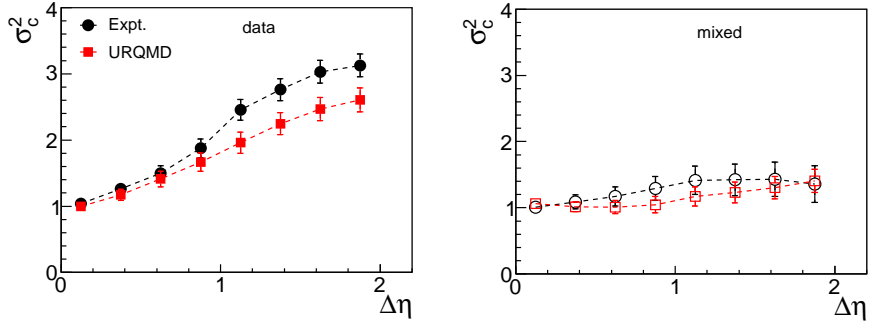,scale=0.6}}
\caption{Dependence of \(\sigma_c^2\) with \et window widths, \(\Delta \eta\) for the experimental, \urq and mixed events.}
\end{figure}

\begin{figure}[bh]
\centerline{\psfig{file=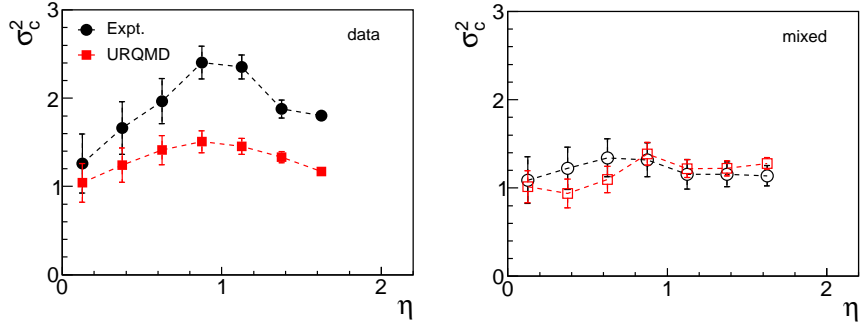,scale=0.6}}
\caption[Fig.7]{\footnotesize Variations of \(\sigma_c^2\) with \et window position for the real, \urq and mixed events. }
\end{figure}

\noindent In order to look for the dependence of \(\sigma_c^2\) on the position of \et windows, two identical \et windows, each of a fixed width, \(\delta\eta\) = 0.5 is chosen and placed adjacent to each other with respect to \(\eta_c\) and the value of \(\sigma_c^2\) is evaluated. The windows are then moved away from \(\eta_c\) in their respective regions in steps of 0.25 \et units until they are  positioned at \(\eta_c \pm\) 1.75. Dependence of \(\sigma_c^2\) on the positions of \et windows, \(\eta\) for the experimental, \urq and mixed events are displayed in Fig.7. It is observed that for the \et window positions at the lowest \(\eta\), \(\sigma_c^2\) \(\sim\) 1. This observation\cite{65}, thus, agrees well with the idea  that clusters produced at around \(\eta_c\) will emit particles in both \ff and \bb regions, inducing  a \lrc that would reduce the value of \(\sigma_c^2\). With increasing separation between the two \et windows, \(\sigma_c^2\) is noticed to first increase upto values 2.4 and 1.6 respectively for the real and \urq data and then it decreases. However, a continuous increase in the \(\sigma_c^2\) values with \(\eta\) position has been reported\cite{13} for $^{16}$O-AgBr and  $^{32}$S-AgBr collisions at 200A GeV/c. The analysis of mixed events give \(\sigma_c^2\) \(\sim\) 1.0 and 1.3, which is slightly larger than expected in the absence of correlations, where \(\sigma_c^2\) is predicted\cite{66,69} to be \(\sim\) 1. Slightly higher values of \(\sigma_c^2\) observed at some \et window positions is due to the fact that \(\langle C\rangle\) \(\ne\) 1, giving \(\sigma_c^2\) \( > \) 1.\\
\begin{figure}[th]
\centerline{\psfig{file=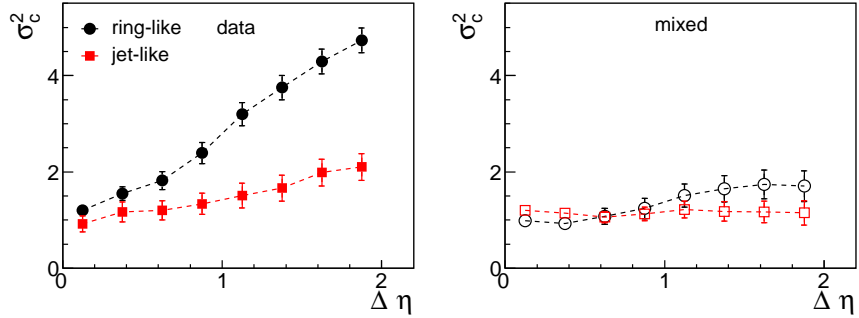,scale=0.6}}
\vspace*{8pt}
\caption{\(\sigma_c^2\) vs \(\Delta\eta\) for the experimental and \urq data compared with the mixed events .}
\end{figure}
\noindent \fb multiplicity \cor in terms of asymmetry variable \(C\) have been investigated mostly at {\footnotesize{RHIC}} energies\cite{65,66,67,68,69}. It has been reported that for $^{197}$Au$^{197}$Au collisions at \(\sqrt{s_{NN}}\) = 200 GeV, quantity, \(\sigma_c^2\) = 2.2 and 3.0 respectively for the central and peripheral collisions for the highest values of \et (position) and \(\Delta\eta\) (width). These investigations also reveal that for 0-20\% central collisions, {\footnotesize{HIJING}} nicely reproduces the values of \(\sigma_c^2\), while \urq overestimates it. Such an overestimation of \(\sigma_c^2\) has been argued that the cluster structure survives in \urq as the hadronic rescattering is not strong enough to destroy it \cite{65} completely. {\footnotesize{AMPT}} model also does not agree with the data and predicts larger \(\sigma_c^2\)  values for the peripheral collisions as compared to those for central ones. Moreover, for very central collisions, the reduction of \(\sigma_c^2\) and hence effective cluster multiplicity, k$_{eff}$ (as \(\sigma_c^2\) \(\sim\) k$_{eff}$ at {\footnotesize{RHIC}} energies is attributed to the idea of cluster melting at these energies\cite{13,65}. The observed maximum values of \(\sigma_c^2\) \(\sim\) 2.4 and 3.0 against \(\eta\) and \(\Delta\eta\) respectively, are larger than that expected from hadron gas model, which gives k$_{eff}$ \(\sim\) 1.5. This results indicates the presence of \srcc, which might arise due to the clusters produced in \ff and or \bb regions and would decay into, on an average,  \(k\) particles. In a detailed study of \fb multiplicity correlations, carried out by UA5 collaboration\cite{19}, the value of k$_{eff}$ has been reported to be \(\sim\) 2. Thus, the large values of \(\sigma_c^2\) observed  in the present study indicates that k$_{eff}$ is stronger than expected from resonance decays as in the case of UA5 findings. The value of large k$_{eff}$ or \(\sigma_c^2\) observed in the present study may be due to (i) large \fll in the number of wounded nucleons in almost inclusive event sample and (ii) non-zero \(\langle C \rangle\) value, which enhances the effect from \src (at least by a value observed for the mixed events) but may be more; Some relevant model calculations are required for a to arrive at a more authentic inference. \\
\begin{figure}[th]
\centerline{\psfig{file=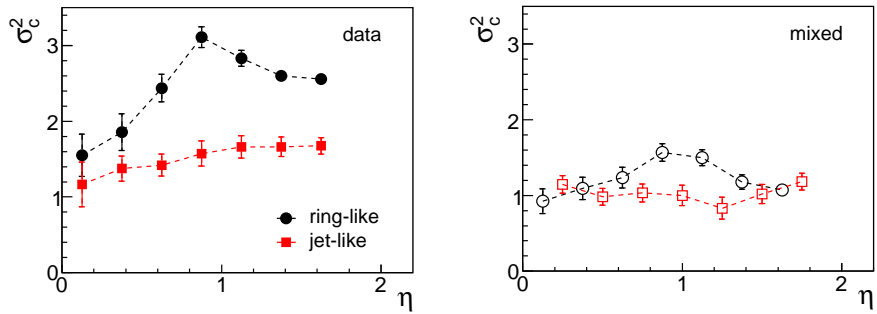,scale=0.6}}
\caption{The same plots as in Fig.7 but for \ring and \jet events. }
\end{figure}

\noindent Shown in Figs.8 \& 9 are the variations of $ \sigma_{c}^2 $ with $ \Delta\eta $ and $ \eta $  for the ring-like and jet-like events. It may be noted from the figure that $ \sigma_{c}^2 $ increases with $ \Delta\eta $ and $ \delta\eta $ in a similar fashion as exhibited by full event sample(Figs 6 and 7). This indicates that the total event sample is dominated by the ring-like events. However, the maximum values of $ \sigma_{c}^2 $ are found to be $ \sim $ 4.3 and 3.1 against  $ \Delta\eta $ and $ \eta $, i.e somewhat larger than those obtained for the entire data sample. For jet like events, although the trends of variations of $ \sigma_{c}^2 $ with  $ \Delta\eta $ and $ \eta $  are the same as those for ring-like or total events, but the magnitude of $ \sigma_{c}^2 $, for a given $ \Delta\eta $ or $ \eta $ is much smaller. The higher values of $ \sigma_{c}^2 $ or $ k_{eff} $ in the case of jet like events as compared to the entire events might be due to large fluctuations in the number of wounded nucleons in the forward and backward moving nuclei. \\
\\

\noindent{\bf 6. Summary  and outlook} \\[-4mm]

\noindent Event-by-event \fll in mean pseudorapidities of relativistic charged particles produced in 11.6A GeV/c $^{197}$Au-AgBr collisions are investigated and the findings are compared with the predictions of relativistic transport model, \urq and model of independent particle emission, {\footnotesize{IPM}}. The findings reveal that the event-wise mean pseudorapidity distribution exhibits a rather longer tail as compared to the reference distribution. \urq simulated data too show similar features but the \fll observed appear to be relatively smaller than the one observed with the real data. Dependence of the \fll measure, \(\Phi_\eta\) on the \et window widths also suggests the presence of \fll probably due to some dynamical origin in the data, which disappear after event mixing.\\

\noindent The study of forward-backward multiplicity \fll is also carried out in terms of the variance (\(\sigma_c^2\)) of the charge asymmetry variable \(C\) and the results are compared with the predictions of \urq and \ipm models. The observed dependence of \(\sigma_c^2\) on the widths and positions of \et windows show that \(\sigma_c^2\) or \(K_{eff}\) increases upto \(\sim\) 3.0, which is significantly higher than those reported by {\footnotesize{UA}}5 collaboration. This indicates that the effect is stronger than that expected due to the presence of \srcc, which might be due to the isotropic decay of cluster-like objects either in forward or backward \et region. The observed higher values of \(\sigma_c^2\) may be attributed to the large \fll in the number of wounded nucleons in almost inclusive event sample. These findings are further confirmed by the results from the  analysis of mixed events, which gives \(\sigma_c^2\) \(\sim\) 1 or somewhat higher, irrespective of the sizes or locations of the \et windows. \urq data too exhibit similar trend of variation of \(\sigma_c^2\) with \(\Delta\eta\) and \et but the magnitudes of the parameter are a bit smaller as compared to one observed with the real data.\\

\noindent Besides the analysis of the total event sample, separate analyses are carried out by identifying the events having \ring and \jet substructures for the charged particle multiplicities. The observed features of \(M_\eta\) distributions for the \ring and \jet events and the \(\Phi_\eta\) dependence on \(\Delta\eta\) for the two types of events indicates that the major contribution to the observed \fll in the data comes from the events with \ring substructures. Moreover, larger values of \(\sigma_c^2\) or \(k_{eff}\) \(\sim\) 4.3 for the \ring events does indicate the presence of much stronger \src in such events as compared to those in the case of \jet events. \\

\noindent It should be mentioned here that asymmetric collisions have been considered in the present study, while the results from {\footnotesize{RHIC}} data involve symmetric collisions. Therefore, a similar study, involving symmetric collisions (may be in future heavy-ion experiment at {\footnotesize{FAIR}} or using the lower energy data from {\footnotesize{RHIC}} and a comparison of the findings with the predictions of the wounded nucleon model might lead to some  interesting conclusions.


\end{document}